\documentstyle[12pt]{article}   

\newcommand{\bea}{\begin{eqnarray}}
\newcommand{\eea}{\end{eqnarray}}

\newcommand{\ba}{\bar a}
\newcommand{\oo}{{\cal O}}
\newcommand{\G}{{\Gamma}}
\newcommand{\ce}{{\cal E}} 
\newcommand{\cp}{{\rm CP}}

\begin{document}

\begin{titlepage}

\begin{flushright}
ITP-SB-97-53 \\
hep-th/9709082 \\
\end{flushright}

\begin{center}
\vskip3em
{\large\bf The Implicit Metric on a Deformation of the 
Atiyah-Hitchin manifold} 

\vskip3em
{Gordon Chalmers\footnote{E-mail:chalmers@insti.physics.sunysb.edu}}\\ 
\vskip .5em
{\it Institute for Theoretical Physics\\ State University of New York\\ 
 Stony Brook, NY 11794-3840, USA\\}
\vskip .4cm
\end{center}

\vfill

\begin{abstract}
\noindent
Using twistor methods we derive a generating function which leads to the 
hyperk\" ahler metric on a deformation of the Atiyah-Hitchin monopole moduli 
space. This deformation was first considered by Dancer through the quotient 
construction and is related to a charge two monopole configuration in a 
completely broken $SU(3)$ gauge theory.  The manifold and
metric are the first members of a family of  hyperk\" ahler manifolds which are
deformations of the
$D_k$ rational singularities of $C^2$.
\end{abstract}

\vfill
\end{titlepage}  

\normalsize

\section{Introduction}\label{intro} The Atiyah-Hitchin metric on the moduli
space of two centered monopoles  in an $SU(2)$ broken gauge theory
\cite{AH} has played a crucial role in considerations of S-duality in field and
string theory as well as in studies of monopole dynamics \cite{sen,sw}.  In
\cite{dancer} Dancer considered a family of deformations of the Atiyah-Hitchin
metric.  He found these deformations by a hyperk\" ahler quotient from the
moduli space of Nahm equations with appropriate bondary conditions.  Although
the hyperk\" ahler quotient constuction is a very powerful tool for establishing
existence and certain properties of hyperk\" ahler manifolds, an explicit
solution for the metric in this approach is in general too difficult.  In this
paper we resort to the second approach to constructing hyperk\" ahler manifolds,
the twistor approach, see e.g. \cite{HKLR}.  In \cite{IR} it  has been shown how
one may use a twistor description to generate certain classes of  metrics
through the generalized Legendre transform \cite{lr}.  The Dancer  manifold is
in the class of manifolds accessible by generalized Legendre  transform and we
utilize this technique to compute the metric implicitly.  

Our main result is the 
generating function used to compute the metric in the generalized Legendre transform:  
$$  
F(z,{\bar z},v,{\bar v},w) =-{1\over 2\pi i} \oint_0 {d\zeta\over \zeta^3}~ \eta(\zeta) 
+ n\oint_C {d\zeta\over\zeta^2} \sqrt{\eta}
$$ 
\bea 
+ {1\over 2\pi i} \oint_{c_+} {d\zeta\over\zeta^2}
(\sqrt{\eta}+{\tilde b})\ln(\sqrt{\eta}+{\tilde b})  
+ {1\over 2\pi i} \oint_{c_-} {d\zeta\over\zeta^2} (\sqrt{\eta}- 
 {\tilde b})\ln(\sqrt{\eta}-{\tilde b})  \ ,
\label{ffunc}
\eea  
where $\eta(\zeta)=z+v\zeta + w\zeta^2-{\bar v}\zeta^3+ {\bar z}\zeta^4$; the contour 
integrals are defined in section 2.  The metrics generated by (\ref{ffunc}) are
also related to the metrics on the moduli space of 
$(2,1)$ monopoles in a maximally broken $SU(3)$ theory \cite{houghton}.  

In section 2, we present the twistor spaces of the families of manifolds which 
are deformations of the Atiyah-Hitchin manifold.  They have a special property which makes
them accessible to the generalized Legendre transform construction, which we describe.
In particular we have to solve a constraint on the parameters of an elliptic curve lying
in the tangent space of $\cp^1$.  In section 3 we show how the constraint 
may be solved implicitly by solving for one of the angle coordinates of the 
parameterization.  In section 4 we give
a description of further metrics  which are deformations of the $D_k$ series.  We further
show how some of these metrics are related to monopole moduli space metrics.

\section{Twistor Spaces for Dancer's family}  
A description of the Atiyah-Hitchin manifold is through the curve 
in $C^3$ 
\bea  
x^2 + y^2 z = 1  \ . 
\label{ahcurve}
\eea   
The metric on which is computed in \cite{AH}. 
Consider next the following family of complex surfaces in $C^3$ 
\bea
x^2 z + (y z + a)^2 = z + a^2 \ .
\label{surface}
\eea  
When $a=0$ we regain the description in (\ref{ahcurve}).  In (\ref{surface}) 
$x, y$ and $z$ are coordinates on $C^3$ and $a$ is a free parameter.  The 
whole family in (\ref{surface}) is related to the ones considered by Dancer
\cite{dancer}.   

In order to put hyperk\" ahler metrics on the set of curves 
in (\ref{surface}) we start with a family of 3-manifolds $T_I$, which will 
serve as twistor spaces for the the hyperk\" ahler metrics.  These twistor 
spaces are constructed from the holomorphic line bundle $\oo (4)$ over $\cp^1$ 
in a manner similar to the construction of the twistor space to the 
Atiyah-Hitchin metric in \cite{AH}.  The defining equation for a 3-manifold 
$T_I$ is: 
\bea 
x^2 (\zeta) \eta (\zeta) + ( y (\zeta) \eta (\zeta) + p (\zeta))^2 =
\eta(\zeta) + p^2 (\zeta) .
\label{tspace}
\eea  
In the last equation (\ref{tspace}), however, the interpretation of 
$x$, $y$, and $\eta$ is different.  Namely, $\zeta$ is the standard coordinate on 
one patch of $\cp_1$ and $\eta(\zeta)$ is a section of the holomorphic line 
bundle $\oo (4)$ (i.e. fourth order polynomial in $\zeta$).  To every 
section $\eta (\zeta)$ we will frequently associate the following elliptic curve
$\ce$ in $\oo (2)$: $\gamma^2 = \eta (\zeta)$.   

To describe the meaning of the coordinates $x (\zeta)$ and $y (\zeta)$ we need 
to consider
the holomorphic tangent bundle $T\cp_1$ of $\cp_1$ covered in a standard way 
by two patches around $\zeta = 0$ and $\zeta = \infty$.  Let $L^m$ be the 
holomorphic line bundle over $T\cp_1$ with transition function 
$e^{- m \kappa /\zeta}$ with $\kappa$ the coordinate on the fiber.  
We also define $L^m(t)$ to be the line bundle $L^m \times \pi^* \oo (t)$ 
where the second term in the product is the pullback bundle from 
$\cp_1$. Then the combinations 
\bea 
\eta_{\pm} (\zeta) =  
  \pm i x (\zeta) \sqrt{\eta (\zeta)} + y (\zeta) \eta (\zeta) + p(\zeta) 
\label{pplusminus}
\eea 
are sections of $L^{\pm m} (2)$.  Finally the deformation is 
given by $p (\zeta) \in \G \oo (2)$; its square is on the same 
footing as $\eta(\zeta)$ and is explicitly parameterized by $p (\zeta) = a + b \zeta
-\ba\zeta^2$ with $b$ real.  

We recall that the reality involution on the sections of holomorphic 
line bundles $\oo(2n)$ is  
\bea  
{\bar\eta}^{(2n)}({\zeta}) = (-1)^n ({\bar\zeta})^{2n}  
 \eta^{(2n)}(-{1/{\bar\zeta}}) \ , 
\eea 
and on the twistor space restricts the section $\eta \in \G \oo (4)$ to depend 
on five real parameters (instead of five complex parameters); it also 
interchanges $p_+$ and $p_-$ in (\ref{pplusminus}).  Moreover the reality 
condition enforces the specific form of the section $p (\zeta)$ above 
with $b \in R, a \in C$.  Thus $I=\{a,b\}$ labels the different 
members of the family of twistor spaces $T_I$.  Restricting to the fibre 
over $\zeta = 0$ we see that the parameter $a$ labels twistor spaces for 
different surfaces (\ref{surface}) (labeled also by $a$) and for fixed $a$, 
the parameter $b$ labels different hyperk\" ahler structures on the same 
surface.

The last ingredient in the description of the twistor spaces $T_I$ is the 
holomorphic 2-form.  We will define it to be
\bea
\omega = d \eta (\zeta) \wedge d \left( \frac{1}{\sqrt{\eta (\zeta)}} 
\ln \frac{\eta_+ (\zeta)}{\eta_- (\zeta)} \right) \ ,  
\eea  
in accordance with \cite{AH} and \cite{dancer}.

As a first step toward recovering the metric we would like to identify a four 
parameter family of twistor lines.  The coefficients of the section 
$\eta (\zeta) \in \G \oo(4)$ provide five parameters and there should be one 
restriction on them coming form the twistor space equation. Indeed, fix 
the coefficients of $\eta (\zeta)$ and consider the zeros of the right hand 
side of (\ref{tspace}).  These correspond to eight points on the elliptic 
curve $\ce$. According to equation (\ref{tspace}) and the meaning of
$\eta_{\pm}$ the divisor of four of the points should correspond to a section of the
holomorphic line bundle $L^{m}$ restricted to the elliptic curve.  The question of how to
split the eight zeros in two groups is governed by the real structure.  The divisor
condition on the second group of zeros is automatically satisfied if the condition on
the divisor of the first group is.  This gives one condition on the coefficients of $\eta
(\zeta)$.  If we fix the four remaining parameters in $\eta (\zeta)$ and vary $\zeta$ we can
recover the sections $\eta_{\pm}$ and this gives us a twistor line.  The solution of the
constraint on the parameters of $\eta (\zeta)$ will ocupy the next section and we close
this section with some additional remarks.

Rational transformations on the $\cp_1$ coordinate 
\bea
\zeta \mapsto \tilde \zeta = \frac{a \zeta + b}{- \bar b \zeta + \bar a}, 
\qquad \vert a \vert^2 + \vert b \vert^2 = 1 ,
\label{trans}
\eea
under which all sections transform covariantly are easily seen to be holomorphic 
maps from the twistor space of a particular surface with given $p (\zeta)$ onto the
twistor space of a surface with $\tilde p (\tilde \zeta) = (\bar b \tilde \zeta + a)^2 p
(\zeta)$.  The combination ${\tilde b}^2 + 4 a \ba$ is invariant under
these transformations and we conclude that it is enough to consider only the
case $a=0, p (\zeta) ={\tilde b} \zeta$ so we have one (real) parameter family of
different hyperk\" ahler metrics, whose twistor spaces we will label as
$T_{\tilde b}$. 

There is still $U(1)$ worth of transformations of type (\ref{trans}) namely 
\bea
\zeta \mapsto \tilde \zeta = e^{i \phi} \zeta  
\label{resisom}
\eea
which leave every $T_{\tilde b}$ invariant.  This isometry is not
tri-holomorphic and we will illustrate it in section 3. 

\section{The Metric and Constraint Equation}

In this section we will discuss how the deformation of the 
Atiyah-Hitchin metric may be found through the generalized Legendre 
transform technique \cite{lr}.  In order to solve for the metric explicitly we need 
to specify a constraint on the twistor lines specified by the parameters 
in $\eta(\zeta)$.  In the derivation of the Atiyah-Hitchin metric using 
the generalized Legendre transform \cite{IR}
this constraint may be solved for in  terms of an overall scale of the
coordinates.  In the case of the  deformation of the Atiyah-Hitchin metric
discussed in this work, there  is an additional scale coming from the
deformation parameter ${\tilde b}$ and  we are unable to determine 
the constraint in this same manner; however,  we can determine it in 
terms of an angular parameter.  
 
Due to the action of the real structure the section $\eta (\zeta)$ has 
the following form
\bea
\eta(\zeta) = z + v \zeta + w \zeta^2 - \bar v \zeta^3 + \bar z \zeta^4 , 
\label{quartic}
\eea
with $z \in C, v \in C$, and $w \in R$.  The 
roots of $\eta (\zeta) + p^2 (\zeta)=0$ then come in complex pairs through 
the real structure and are denoted 
$\alpha, \beta, -1/{\bar \alpha}, -1/{\bar \beta}$.  The
following four points on the elliptic surface $\ce$, 
$$
\ce: \; (\alpha, \sqrt{\eta (\alpha)}), (\beta, \sqrt{\eta (\beta)}), 
$$
\bea 
\qquad\qquad
(-1/{\bar \alpha},- \sqrt{\eta (-1/{\bar \alpha})}), 
  (-1/{\bar \beta}, - \sqrt{\eta (-1/{\bar \beta})}) \ ,
\eea
then are the zeros of the section $\eta_+$. The condition on the 
divisor then could be written as
\bea
\left( \int_{\alpha}^{\infty} + \int_{\beta}^{\infty} - 
\int_{-1/{\bar \alpha}}^{\infty} - \int_{-1/{\bar \beta}}^{\infty} \right) 
 \frac{d\zeta}{\sqrt{\eta (\zeta)}} = 2 .  
\label{firstcon}
\eea
The integral representation in (\ref{firstcon}) is defined by an analytic 
continuation.  There is a square root branch cut of the integrand which 
we take to run between the zeros $\alpha$ and $-1/{\bar\beta}$ (and 
$-1/{\bar\alpha}$ and $\beta$).  Due to the branch cut the ambiguity is 
given by integrals along these pairs of zeros; they give complete 
elliptic integral contributions to (\ref{firstcon}).  This ambiguity 
will be fixed by comparing the metric to its known asymptotic form.  

Alternatively, the constraint may be expressed as the derivative 
$F_w=0$ of the function 
$$  
F(z,{\bar z},v,{\bar v},w) =-{1\over 2\pi i} \oint_0 {d\zeta\over \zeta^3}~ \eta(\zeta) 
+ n\oint_C {d\zeta\over\zeta^2} \sqrt{\eta} 
$$ 
\bea 
+ {1\over 2\pi i} \oint_{c_+} {d\zeta\over\zeta^2}
(\sqrt{\eta}+{\tilde b})\ln(\sqrt{\eta}+{\tilde b})  
+ {1\over 2\pi i} \oint_{c_-} {d\zeta\over\zeta^2} (\sqrt{\eta}- 
 {\tilde b})\ln(\sqrt{\eta}-{\tilde b})  
\label{fcon}
\eea 
The first term in (\ref{fcon}) is given by a contour integral around the origin and 
the second one $C$ is defined by an integral around one of two branch cuts running 
between the pairs of zeros ($\alpha_0$,$-1/{\bar\beta}_0$) and ($-1/{\bar\alpha}_0,\beta_0$) 
to the equation $\eta(\zeta)=0$.  The remaining two contours $c_\pm$ are 
specified by the zeros of the equation $\eta(\zeta)+{\tilde b}^2 \zeta^2=0$.    
Explicitly one may write them as line integrals on a sheet 
of the double cover of the complex plane specified by branch cuts 
running through the between the zeros of $\eta(\zeta)+{\tilde b}^2 \zeta^2=0$.  If 
${\tilde b}=0$ then these integrals would be complete elliptic integrals.  The 
representation of the twistor space constraint in (\ref{fcon}) defines the generating 
function used for computing the metric in the Legendre
transform technique.  By construction it satisfies 
\bea  
{\partial\over\partial{w_a}} {\partial\over\partial {w_b}} F=  
  {\partial\over\partial {w_{(a+c)}}} {\partial\over\partial{w_{(b-c)}}} F  \ ,
\eea   
where the coordinates $w_j$ are defined through $\eta^{(4)}(\zeta)= 
 \sum_{j=0}^4 w_j \zeta^j$

In order to solve the constraint (\ref{fcon}), we are faced with the 
problem of adding together two incomplete elliptic integrals of 
first kind.  Unfortunately we have not been able to solve the constraint 
equation explicitly for a scale parameter (for example, the variable $w$) 
as is done in finding the Atiyah-Hitchin metric in this formulation \cite{IR}.  

We take a rational transformation which sends one of the roots of 
(\ref{quartic}) to infinity and then the transformed quartic is 
\bea 
{\tilde b}^2 \left[ -a{\bar b} \zeta^2 + (a{\bar a}-b{\bar b}) \zeta  
 +b{\bar a}\right]^2 = r_1 \zeta^3- r_2 \zeta^2  - r_1\zeta \ .
\label{tqcon} 
\eea 
Here $r_1$ and $r_2$ are both real numbers.  The zeros of 
(\ref{tqcon}) are $x_1$, $x_2$ and $-1/{\bar x}_1$, $-1/{\bar x}_2$.  
The limit ${\tilde b}\rightarrow 0$, which generates the Atiyah-Hitchin 
curve, is subtle because in this case two of the roots tend to 
zero and infinity.  In (\ref{tqcon}) the five parameters given by the
coefficients of the original section $\eta (\zeta)$ in (\ref{quartic}) have been
traded in exchange for the three angles describing the $\zeta$ rotation and
$r_1$ and $r_2$.  

The constraint on the twistor space is alternatively found 
by adding together the elliptic integrals in (\ref{fcon}) directly.  
Using the identity (\ref{identity}) after inverting the 
integrals in (\ref{fcon}) we obtain a second form 
\bea
x - \frac{r_2}{3 r_1} = \wp ({m \sqrt{r_1}}/{2};g_2,g_3) \ , 
\label{partcon}
\eea  
where $x$ is the root of the cubic equation in (\ref{tqcon}) which 
goes to infinity when $r_1\rightarrow 0$ and $\wp$ is the Weierstrauss 
function.  The modular parameters are 
\bea 
g_2 = 4\Bigl\{ 1+ 3({r_2\over 3r_1})^2\Bigr\} 
\eea 
and
\bea 
g_3 = 4 {r_2\over 3r_1} \Bigl\{ 1+ 2({r_2\over 3r_1})^2\Bigr\}  \ .  
\label{modular}
\eea  
The equation (\ref{partcon}), when regarded as giving $r_1$, $r_2$, or 
the ratio $r_2/r_1$ in (\ref{partcon}), is transcendental and may not 
be solved for explicitly.  However, in principle we may solve for one 
of the angular coordinates appearing in $x$ in (\ref{partcon}) through 
the equation (\ref{tqcon}); this form is useful for a numerical 
implementation of describing the metric.   In the appendix we express the 
function in (\ref{fcon}) in terms of the rotated coordinates.  

\subsection{$U(1)$ Isometries}  

In the presence of isometries the construction of the metrics using 
the generalized Legendre transform may be simplified.  In this section we
describe how a $U(1)$ isometry, both tri-holomorphic and non-tri-holomorphic, 
demonstrates this.  In the process we 
will derive a theorem due to Boyer and Finley \cite{bf}.  

In the case of a tri-holomorphic isometry one may always 
choose a set of complex coordinates so that for a 
vector field $\psi$ and Kahler form $K$ \cite{HKLR},
\bea 
\xi=(\partial_u - \partial_{\bar u}) \quad\quad \xi K=0 . 
\label{triisom}
\eea 
Furthermore, if the isometry is only a generic $U(1)$ 
we may find coordinates $u'$ and ${\bar u}'$ such that 
\bea 
\psi= (u'\partial_{u'} - {\bar u}'\partial_{{\bar u}'}) \quad\quad 
 \psi K=0 \ , 
\label{genisom}
\eea 
or with a parameterization $u=e^{i\theta} r$ 
\bea  
\partial_\theta K=0 \ .
\eea 
The former case is special in that metrics possessing 
a tri-holomorphic $U(1)$ isometry are {\it always} constructed 
locally from an ${\cal O}(2)$ coordinate:  The Kahler potential 
always satisfies (\ref{triisom}) when written as a function 
of $u+{\bar u}$.  This case, however, is not relevant to our study as both
the Atiyah-Hitchin metric and its deformation possess a generic $U(1)$ isometry 
which is not tri-holomorphic (unless one is in the asymptotic regime).  In 
both cases an ${\cal O}(4)$ section is used to formulate the metrics. 

First we demonstrate the $U(1)$ action explicitly on the 
generating function.  Recall that the Atiyah-Hitchin metric is 
found in the generalized Legendre transform from  \cite{IR}
\bea 
F_{AH} = -{m\over 2\pi i} \oint_0 {d\zeta\over \zeta^2} \eta(\zeta) 
 + \oint_C {d\zeta\over \zeta} \sqrt{\eta(\zeta)} \ , 
\label{ahgen}
\eea 
where the contour $C$ encloses all four roots of $\eta$ as described 
in \cite{IR} (or as given in \ref{fcon}).  

Consider a transformation of the coordinate $\eta(\zeta)$ 
in (\ref{quartic}) used to describe these metrics  
given by a constant phase shift $\zeta\rightarrow e^{i\alpha}\zeta$.  Our 
section transforms as  
\bea 
\eta\rightarrow e^{2i\alpha} \Bigl( e^{-2i\alpha} z + 
 e^{-i\alpha} v \zeta +  w \zeta^2  -  e^{i\alpha} {\bar v} \zeta^3 
 + e^{2i\alpha} {\bar z} \zeta^4 \Bigr) \ .
\eea 
Upon redefining our coordinates as $z'=e^{-2i\alpha} z$ and 
$v'=e^{-i\alpha} v$ we see explicitly how the generating 
function remains unchanged for the Atiyah-Hitchin metric.  The 
invariance of the generating function for the Atiyah-Hitchin 
metric $F_{\rm AH}$ is due to its particular functional form:  The 
measure of the integrand transforms in the opposite manner and 
eliminates the phase $e^{i\alpha}$ from appearing explicitly.  Of 
course, the roots to $\eta(\zeta)+b^2\zeta^2=0$ change.    

The example described in (\ref{fcon}) also possesses this $U(1)$ invariance 
provided the deformation parameter also gets rescaled. 
This is the isometry listed in (\ref{resisom}).  In general this invariance 
exists in further examples of metrics provided that the $F$-functions 
in such cases remains unchanged. 

The invariance under these phase shifts explicitly realizes a $U(1)$ 
isometry: under this rotation we may choose our coefficient $v$ 
to be real.  The Kahler potential is found from the Legendre transform 
\bea
K(z,{\bar z},u,{\bar u}) = F(z,{\bar z},v,{\bar v},w)- uv-{\bar u}{\bar v}  
\qquad u=F_v
\label{transform}
\eea 
after taking into account the constraint $F_w=0$.  The components of the 
metric are found by appropriate derivatives of the Kahler potential 
($ds^2=\partial_z \partial_{\bar z} K(z,{\bar z},u,{\bar u}) ~dz\otimes d{\bar z} 
+ \ldots$). 

This reality property of $u$ after using the phase invariance of $F$ 
may be used to explicitly simplify the construction of the Kahler
potential in (\ref{transform}).  We define the coordinates
\bea 
v=J e^{i\theta} \quad\quad u=re^{i\theta} \ .
\eea 
By using the isometry we can reduce the Legendre transform, which 
usually requires two parameters, to only one  
\bea 
K(z,{\bar z}, r) = F(z,{\bar z},J) - J r  \quad\quad 
 r=2 \partial_J F(z,{\bar z},J) \ . 
\label{simptrans}
\eea 
The Kahler potential is independent of $\theta$ and 
we obtain the previous result of Boyer and Finley (\ref{genisom}), which 
is essentially (\ref{simptrans}). 

\section{$(2,1)$ Charge Monopole Moduli Space}

Although we are unable to compute the metric explicitly from the 
generating function, in the following we note its relations to 
the metric on the $(2,1)$ centered moduli space occuring in a completely 
broken $SU(3)$ gauge theory recently considered in \cite{houghton}.  

The asymptotic form of the metric generated by
(\ref{fcon}) is found by taking the limit $\eta(\zeta)\rightarrow
{\tilde\eta}^2(\zeta)$, where ${\tilde\eta}^2(\zeta)=z+x\zeta-{\bar z}\zeta^2$ 
(i.e., a coordinate of the $O(2)$ type).  This limit has been discussed 
in both \cite{IR,ch} and we do not repeat the analysis here.  In 
the asymptotic regime the metric is (with $m=1$)
$$   
ds^2=(1-{n\over \vert{\vec r}\vert} +{1\over \vert {\vec r}-{\vec \lambda}\vert}+ 
 {1\over \vert {\vec r}+{\vec \lambda}\vert}) ~d{\vec r}\cdot d{\vec r} 
$$
\bea
+ (1-{n\over \vert{\vec r}\vert}+{1\over \vert {\vec r}-{\vec \lambda}\vert}+ 
 {1\over \vert {\vec r}+{\vec \lambda}\vert})^{-1}  
 d{\tilde\phi} d{\tilde\phi} \ .   
\label{nut} 
\eea  
where ${\vec\lambda}=({1\over{\sqrt 2}}{\rm Re}a, 
{1\over{\sqrt 2}}{\rm Im}a, b)$.  The frame $d{\tilde\phi}$ is 
given by $d{\tilde\phi}=d\xi+{\vec w}\cdot {d\vec r}$ with 
\bea 
{\vec\nabla}\times {\vec w}= -{\vec\nabla} \Bigl({n\over \vert{\vec r}\vert}- 
{1\over \vert {\vec r}-{\vec \lambda}\vert}-{1\over \vert {\vec r}+{\vec \lambda}\vert} 
\Bigr) \ .
\eea 
In the preceeding section we have used an $SO(3)$ rotation to set $a={\bar a}=0$:  It 
is shown in \cite{dancer} that these rotations act on the spaces 
$M({\vec\lambda})$ by rotating the perturbation, $M(R\cdot{\vec\lambda})$.    
The form (\ref{nut}) is the asymptotic form of the (double cover) of the 
Atiyah-Hitchin metric once we set ${\vec \lambda}=0$.  The general perturbation 
we have in (\ref{nut}) is related to the form of the multi-center 
Taub-NUT class of metrics \cite{tnmetrics}.  

In fact, if we replaced the generating function 
in (\ref{fcon}) with a summation over $k$ distinct points $b_j$ as in 
$$  
F^{(k)} (z,{\bar z},v,{\bar v},w) = - {1\over 2\pi i} \oint_0  
 {d\zeta\over \zeta^3}~ \eta(\zeta) 
+ n\oint {d\zeta\over\zeta^2} \sqrt{z}(\zeta)  
$$
$$
+ \sum_{j=1}^k {1\over 2\pi i} \oint_{c_{+,j}} {d\zeta\over\zeta^2}
(\sqrt{z}(\zeta)+b)\ln(\sqrt{z}(\zeta)+b)  
$$
\bea
+ \sum_{j=1}^k {1\over 2\pi i} \oint_{c_{-,j}} {d\zeta\over\zeta^2}
(\sqrt{z}(\zeta)-b)\ln(\sqrt{z}(\zeta)-b)  \ ,
\label{multifcon} 
\eea 
then we would obtain the asymptotic form of a $2k+1$-multi-center 
Taub-NUT metric and is symmetric under ${\vec r}=-{\vec r}$.   The asymptotic 
form of the monopole moduli space metrics with charges $(2,1,\ldots,1)$ 
in the center-of-mass of the $(2,)$ charge has the form (\ref{multifcon}).  The 
metrics in (\ref{multifcon}) are the NUT-perturbations of the $D_k$ space of 
metrics for $n=0$ (the $D_k$ spaces of metrics within the generalized Legendre 
transform were described in \cite{IR}); the first member $k=1$ of which is related 
to the deformation of the Atiyah-Hitchin metric after a discrete quotient by a $Z_2$.   

In \cite{houghton} it was shown using Nahm's equations 
that the Dancer's family of metrics is the same as that describing 
the $SU(3)\rightarrow U(1)^2$ monopole monopole moduli space with
charges $(2,1)$ after taking the infinite mass limit of the
$(,1)$ charge.  The three parameters $(a,{\bar a},b)$ labelling the
deformation is now the position of the fixed $(,1)$ charge.  In 
our derivation we have used an $SO(3)$ rotation to send $a={\bar a}=0$ 
and expressed the metric in the center of mass of the $(2,)$ charges.  
We find that by 
comparing the asymptotic behavior of the ``fixed" monopole moduli 
space metric considered by Houghton that the Dancer's metric is  
determined by specifying $n=1$ in (\ref{fcon}).

Finally, we relate the construction above to generate implicitly 
the metrics on the moduli spaces of the $(2,1)$ monopoles.  The 
parameters $\vec\lambda$ label both the perturbation in the Dancer's 
metrics or alternatively the positions of the fixed $(,1)$ charge.  
As noted in \cite{ch} if we allow this position to vary then 
in principle we may also construct the $(2,1)$ metric by promoting 
the parameters in $p(\rho)=a+b\rho-{\bar a}\rho^2$ to coordinates.  
In this case the generating functions described above would give 
the metric on the $8$-dimensional space which corresponds to this 
moduli space metric in the center of mass of the $(2,)$ charge.  
The moduli space metrics described in \ref{multifcon} follow from 
the construnction using Nahm's equations by \cite{houghton} and 
prove the form conjectured in \cite{ch}.

\section{Discussion}  

We have described the deformation of the Atiyah-Hitchin metric as 
well as the NUT-version of the ALE $D_1$ space implicitly 
through the generalized Legendre transform technique.  These metrics 
are of interest for both their relations to problems in monopole 
dynamics and duality in field theory.  

It would be interesting to further explore the detailed form of the 
deformation of the Atiyah-Hitchin metric which was implicitly described 
in this work.  It has been used to describe the moduli space of vacua 
for an $SU(2)$ $N=4$ supersymmetric gauge theory in three dimensions with one 
hypermultiplet.  Such moduli spaces are related by dimensional reduction 
to $N=2$ gauge theory moduli spaces in four dimensions which possess singularities 
describing massless mutually non-local dyons.  

Last, the Legendre transform technique has been quite useful 
in describing complete hyperk\" ahler metrics; it is known that 
all $4k$-dimensional metrics with $k$ tri-holomorphic isometries 
may be generated in this formulation.  There
remains the question on whether the generalized Legendre transform 
is as extensive in the classification of hyperk\"ahler metrics 
not possessing the tri-holomorphic isometries.  The 
monopole moduli space metrics do not have these isometries 
and it would be interesting to find their construction implicitly 
at the level of the generating function.     

Further work in the construction of the moduli space metrics is in 
progress.  One natural outcome is in the construction of all the 
ALE space metrics as found in the twistor description described 
in \cite{AH}.  One example of such a construction is in the 
higher dimensional analogs of the $A_k$ series, which have recently 
appeared in the form of the $(1,1,\ldots,1)$ charge metrics in the 
higher rank gauge groups.  However,another example consists 
in thinking of the $D_2$ ALE metric as removing the "NUT" addition 
to the generating function describing the Atiyah-Hitchin metric.   Further 
examples follow from the higher charge monopole modulit space metrics.  
In addition, most remaining ALE spaces (and their higher dimensional 
analogs) have natural embeddings in the work, for example, recently performed 
on symmetric configurations of monopoles \cite{symm}.

\vskip .2in
\noindent{\large\bf Acknowledgements}  
\vskip 1em

\noindent{G.C. thanks I.T.\ Ivanov for collaboration and Martin Rocek for 
discussions.  This work was supported in part by  NSF grant No.~PHY 9722101.}

\vskip .3in
\section{Appendix}  
\vskip .2in 

The addition formula for a quartic of general type as in 
(\ref{quartic}) is very cumbersome and we will use a rational transformation
(\ref{trans}) on $\zeta$ to put the quartic into a simpler form.  

We rotate the quartic (\ref{firstcon}) into Weierstrauss form 
by using the $SL(2,C)$ transformation in (\ref{trans}), 
under which we have  
\bea  
\eta(\zeta)\rightarrow {1\over (-{\bar b}\zeta+{\bar a})^4} 
 (r_1 \zeta^3-r_2\zeta^2-r_1 \zeta) 
\qquad 
d\zeta \rightarrow {d\zeta\over (-{\bar b}\zeta+{\bar a})^2} \ , 
\eea 
and 
\bea  
p(\zeta)\rightarrow {{\tilde b}\over (-{\bar b}\zeta+{\bar a})^2} 
 (a\zeta+b)(-{\bar b}\zeta+{\bar a}) \ . 
\eea 
The quartic $\eta(\zeta)-{\tilde b}^2\zeta^2=0$ is then placed in the 
form (\ref{tqcon}).

Upon shifting the integrand by 
$\zeta \mapsto \zeta - r_2/{3 r_1}$ we obtain the constraint in 
the form 
\bea  
z_1+z_2-{\tilde z}_1-{\tilde z}_2 = {m\over 2} \sqrt{r_1} 
\label{zs}
\eea 
where 
\bea  
z_1=\wp^{-1}(x_1-{r_2\over 3r_1}) \quad\quad {\tilde z}_1= 
\wp^{-1}(-{1\over{\bar x}_1}-{r_2\over 3r_1})  \ . 
\label{zeq}
\eea 
The values for $z_2$ are give by replacing $x_1$ with $x_2$ in 
(\ref{zeq}).  In the above we have used the definition of the 
Weierstrauss $\wp$-function, 
\bea  
z=\int_{\wp(z)}^\infty {d\zeta\over (4\zeta^3-g_2 \zeta-g_3)^{1/2}} \ .
\eea 
The modular parameters of the Weierstrauss 
$\wp$-functions in (\ref{zeq}) are given in (\ref{modular}). 

In the following we shall moderately simplify the form 
in (\ref{zs}).  Due to the reality condition we have the 
immediate identity 
\bea 
z_1+z_2 = -{\tilde z}_1-{\tilde z}_2 = m\sqrt{r_1}  \ . 
\label{relation}
\eea 
The value of $z_1$ may be determined from ${\tilde z}_1$ through 
(\ref{zeq}) and (\ref{relation}).  The standard
addition formula of $\wp$-functions gives 
\bea  
\wp(z_1+z_2)=\wp(z_1)+\wp(z_2) +{1\over 4}  
 \Bigl[ {\wp'(z_1)-\wp'(z_2)\over \wp(z_1)-\wp(z_2)}\Bigr]^2  \ . 
\label{identity} 
\eea 
Using this identity and the defining differential 
equation of the $\wp$-function, 
\bea  
\wp'^2(z) = 4\wp^3(z) - g_2 \wp(z) - g_3 \ , 
\eea 
we further obtain the form of the constraint in the form  
\bea 
\wp(m{\sqrt r_1};g_2,g_3) = -(x_1+x_2) + {2r_2\over 3r_1} 
 + {{\tilde b}^2\over r_1} \Bigl\{ -a{\bar b}(x_1+x_2) + a{\bar a} 
- b{\bar b} \Bigr\}^2  \ ,  
\label{geq}
\eea 
in terms of two roots to (\ref{tqcon}) which are not conjugates 
to one another (i.e. $x_1\neq -1/{\bar x}_2$).  

We end the appendix with some formulas expressing the generating 
function in terms of the coordinates $r_1$,$r_2$ and $a$,$b$.  The 
first term in (\ref{fcon}) is 
$$ 
F_1 = {1\over 2\pi i} \oint {d\zeta\over \zeta^2}~ {\sqrt \eta}(\zeta)  
 \ln({\sqrt \eta}\pm b)   
$$
\bea 
 = {{\sqrt r_1}\over 2(a{\bar b})^2} \int_{z_1}^{-1/{\bar z}_1} dz ~{\wp'(z)^2\over 
 \Bigl[(\wp(z)+{r_2\over 3r_1}+{b\over a}) 
 (\wp(z)+{r_2\over 3r_1}-{{\bar a}\over {\bar b}}) \Bigr]^2 } \ , 
\label{contone}
\eea 
evaluated between the zeros of $\sqrt{z}\pm b=0$.  The form 
(\ref{contone}) rewritten into, 
\bea  
F_1 = {{\sqrt r_1}\over 2} \int_{z_1}^{-1/{\bar z}_1} dz ~\Bigl\{ {d\over dz} \ln\Bigl( {
 \wp(z)+{r_2\over 3r_1}+{b\over a} \over  
 \wp(z)+{r_2\over 3r_1}-{{\bar a}\over {\bar b}}} \Bigr) \Bigr\}^2  \ .
\eea 
The second contribution is a total derivative, 
$$
F_2 = {1\over 2\pi i} \oint {d\zeta\over\zeta^2}  
  ~b\ln(\sqrt{\eta}\pm b)  \ ,
$$ 
which after the $SL(2,C)$ rotation is 
\bea 
F_2 = -{{\tilde b}\over a{\bar b}} \int_{z_1}^{-1/{\bar z}_1} ~dz {\wp'(z)\over  
 \Bigl[(\wp(z)+{r_2\over 3r_1}+{b\over a}) 
 (\wp(z)+{r_2\over 3r_1}-{{\bar a}\over {\bar b}}) \Bigr]} \ . 
\label{second} 
\eea 
The integral (\ref{second}) may be expressed as a total 
derivative.  The result is 
\bea 
F_2 = - {\tilde b} \ln \Bigl[ { \wp(z)+{r_2\over 3r_1}+{b\over a}\over 
 \wp(z)+{r_2\over 3r_1}-{{\bar a}\over {\bar b}} } \Bigr] \Bigl\vert_{z_1}^{-1/{\bar z}_1} \ , 
\eea 
evaluated at the end-points of the integral in (\ref{second}).  
These evaluations of the generating function will be useful when 
constructing the metric using the formulas \cite{lr} and a possible 
numerical evaluation.

\end{document}